\documentclass[twocolumn,pra,showpacs]{revtex4}
\pdfoutput=1
\usepackage{setspace}
\usepackage{amsfonts,amsmath,amssymb}
\usepackage{txfonts}
\usepackage{graphicx, color}
\usepackage{hyperref}

\begin{document}

\title{Nonequilibrium dynamics of vortex arrest in a finite-temperature Bose-Einstein condensate}

\author{T. M. Wright}
\affiliation{Jack Dodd Centre for Quantum Technology, Department of Physics, University of Otago, PO Box 56, Dunedin, New Zealand}

\author{A. S. Bradley}
\affiliation{Jack Dodd Centre for Quantum Technology, Department of Physics, University of Otago, PO Box 56, Dunedin, New Zealand}

\author{R. J. Ballagh}
\affiliation{Jack Dodd Centre for Quantum Technology, Department of Physics, University of Otago, PO Box 56, Dunedin, New Zealand}

\begin{abstract}
We perform finite-temperature dynamical simulations of the arrest of a rotating Bose-Einstein condensate by a fixed trap anisotropy, using a Hamiltonian classical-field method.  We consider a quasi-two-dimensional condensate containing a single vortex in equilibrium with a rotating thermal cloud.  Introducing an elliptical deformation of the trapping potential leads to the loss of angular momentum from the system.  We identify the condensate and the complementary thermal component of the nonequilibrium field, and compare the evolution of their angular momenta and angular velocities.  By varying the trap anisotropy we alter the relative efficiencies of the vortex-cloud and cloud-trap coupling.  For strong trap anisotropies the angular momentum of the thermal cloud may be entirely depleted before the vortex begins to decay.  For weak trap anisotropies, the thermal cloud exhibits a long-lived steady state in which it rotates at an intermediate angular velocity.
\end{abstract}


\date{\today}

\maketitle
\section{Introduction}\label{sec:Introduction}
Since the first experimental realizations \cite{Anderson95,Davis95,Bradley95} of dilute-gas Bose-Einstein condensates, there has been much interest in the properties of quantum vortices in such systems, and the effect of thermal atoms on their dynamics and stability.  The description of the dynamics of vortices in the presence of thermal excitations provides a challenging test for dynamical theories of cold bosonic gases, and promises new insights into the role of thermal excitations in the dynamics of vortices in systems less amenable to such \emph{ab initio} descriptions \cite{Sonin97,Thouless96,Donnelly91,Blatter94}.
The stability of a vortex state in a dilute condensate was first investigated theoretically by Rokhsar \cite{Rokhsar97}, who showed that a vortex is subject to decay in the presence of a nonrotating thermal cloud, providing a physical interpretation of a negative-energy excitation of the vortex previously found by Dodd \emph{et al.} \cite{Dodd97}.  Fedichev and Shlyapnikov \cite{Fedichev99} then put forward an analytic theory of the dynamics of vortex decay in the presence of a nonrotating thermal cloud, based on a two-fluid model of superfluidity.  Madison \emph{et al.} \cite{Madison00} subsequently observed experimentally a nonexponential decay in vortex survival probability, and the first evidence of an increase in the displacement of the vortex from the trap center during the decay, which was also observed in the experiments of Anderson \emph{et al.} \cite{Anderson00}.  Zhuravlev and co-workers \cite{Zhuravlev01} provided an analytic description of the decay of vortex arrays, building on the work of \cite{Fedichev99}, and including the rotational dynamics of a nonstationary thermal cloud on the basis of the theory of Gu\'ery-Odelin \cite{Guery-Odelin00}.  Their work predicted two limiting regimes of relaxation dynamics, depending on the relative efficiencies of vortex-cloud and cloud-trap coupling: the so-called \emph{rotating trap} limit, in which the vortex array and thermal cloud relax collectively as a single rigid body, and the \emph{static trap} limit, in which the thermal cloud is quickly arrested by the trap anisotropy, and nonexponential decay \cite{Fedichev02} of the array rotation follows.   

Abo-Shaeer \emph{et al.} \cite{Abo-Shaeer02} performed experiments in the rotating trap regime, and observed the expected exponential decay and strong temperature dependence of the decay rate.  Rosenbusch \emph{et al.} \cite{Rosenbusch02} observed the decay of a single vortex experimentally, and found a much less severe dependence on temperature than that of \cite{Abo-Shaeer02}.  They conjectured that the thermal cloud was rapidly arrested by residual trap anisotropy, leading to the static-cloud vortex-decay scenario of \cite{Fedichev99}.  Classical-field simulations performed by Schmidt \emph{et al.} \cite{Schmidt03} focused on the dynamics of a strongly nonequilibrium `phase-imprinted' vortex state.  Duine \emph{et al.} \cite{Duine04} presented an analytical description of the decay of a vortex in a nonrotating thermal cloud, obtained from the stochastic field theory of \cite{Stoof99} using a variational ansatz \cite{Duine01}, which was subsequently extended to include the effects of cloud rotation by Bradley and Gardiner \cite{Bradley05a}.

In the present paper we consider a condensate containing a vortex initially at thermal and rotational equilibrium with a rotating thermal cloud in a highly oblate trap, which is isotropic in the plane.  Such a configuration is obtained as an ergodic classical-field equilibrium with fixed angular momentum on the order of $\hbar$ per atom about the trap axis \cite{Wright09}.  Due to the conservation of angular momentum, this rotating equilibrium configuration is stable, provided that the trapping potential remains invariant under rotations about its axis.  We then introduce an elliptical deformation of the trap which breaks this rotational symmetry, leading to the loss of angular momentum from the atomic field and thus the slowing of the rotating cloud, and consequently the decay of the vortex.  Our simulations are the first to describe the arrest of the rotation of both the condensed and noncondensed components of the field following the introduction of a trap anisotropy, and our method describes the coupled relaxation dynamics of the two components implicitly.  As predicted by \cite{Zhuravlev01}, we find that the response of the condensed and noncondensed components of the field may be different depending on the relative efficiencies of vortex-cloud and cloud-trap coupling, where the latter depends on the ellipticity of the trap deformation.  By varying the anisotropy over a range of values and monitoring the evolution of the condensed and noncondensed components of the field, we observe a rich phenomenology, ranging from an adiabatic steady state, to violently nonequilibrium dynamics, in which the rotation of the thermal cloud essentially decouples from that of the condensate.  

This paper is organized as follows.  In Sec.~\ref{sec:Formalism} we discuss our classical-field formalism, and the parameters of the physical system we simulate.  In Sec.~\ref{sec:Results} we discuss the results of a representative simulation.  In Sec.~\ref{sec:Anisotropy_dependence} we discuss the dependence of the dynamics of the vortex and thermal field on the trap anisotropy.  In Sec.~\ref{sec:Conclusions} we summarize our findings and present our conclusions.

\section{Formalism}\label{sec:Formalism}
\subsection{Classical-field method}
The general formalism of (projected) classical-field methods has recently been reviewed at length in \cite{Blakie08}, and so here we merely outline the particular method we use in this work.  The system we represent is described by the second-quantized, cold-collision ($s$-wave) Hamiltonian
\begin{equation}\label{eq:FullH}
H=\int d\mathbf{x}\; \hat{\Psi}^\dag(\mathbf{x}) H_{\mathrm{sp}} \hat{\Psi}(\mathbf{x}) +\frac{U}{2} \int d\mathbf{x}\;\hat{\Psi}^\dag(\mathbf{x})  \hat{\Psi}^\dag(\mathbf{x}) \hat{\Psi}(\mathbf{x})\hat{\Psi}(\mathbf{x}), 
\end{equation}
where $\hat{\Psi}(\mathbf{x})$ is the bosonic field operator, which satisfies $[\hat{\Psi}(\mathbf{x}),\hat{\Psi}^\dagger(\mathbf{x}')]=\delta(\mathbf{x}-\mathbf{x}')$, and $U=4\pi\hbar^2a/m$ is the interaction potential written in terms of scattering length $a$ and atomic mass $m$.  The single-particle Hamiltonian is of form
\begin{equation}
	H_\mathrm{sp} = \frac{-\hbar^2\nabla^2}{2m} + V_0(\mathbf{x}) + V_\epsilon(\mathbf{x}) \equiv H^0_\mathrm{sp} + V_\epsilon(\mathbf{x}),
\end{equation}
where $V_0(\mathbf{x})=(m/2)\{\omega_r^2(x^2+y^2)+\omega_z^2z^2\}$ is a cylindrically symmetric trapping potential and $V_\epsilon(\mathbf{x})=\epsilon m\omega_r^2(y^2-x^2)$ is an additional anisotropic potential.  In deriving the classical-field method, we replace the field operator $\hat{\Psi}(\mathbf{x})$ with the projected classical field $\psi(\mathbf{x})=\sum_{n\in\mathbf{L}}\alpha_n \phi_n(\mathbf{x})$.  The sum here is over eigenmodes of the cylindrically symmetric single-particle Hamiltonian [$H_\mathrm{sp}^0\phi_n(\mathbf{x})=\epsilon_n\phi_n(\mathbf{x})$] with eigenvalues satisfying $\epsilon_n\leq E_R$, where $E_R$ is the \emph{cutoff energy}.  The projected classical field is thus a vector in a low-energy subspace ($\mathbf{L}$) of the appropriate single-particle Hilbert space.  Making this replacement we obtain the classical Hamiltonian
\begin{equation}\label{eq:HCF}
H_{\mathrm{CF}}=\int d\mathbf{x}\; \psi^*(\mathbf{x}) H_{\mathrm{sp}}
\psi(\mathbf{x}) +\frac{U}{2} |\psi(\mathbf{x})|^4.
\end{equation}
Defining the projector
\begin{equation}\label{eq:Pdef} 
{\cal P}f(\mathbf{x})\equiv\sum_{n \in
\mathbf{L}}\phi_n(\mathbf{x})\int d\mathbf{y}\;
\phi_n^*(\mathbf{y})f(\mathbf{y}),  
\end{equation} 
we can express the Hamilton's equation for $\psi(\mathbf{x})$ obtained from Eq.~(\ref{eq:HCF}) as
\begin{equation}\label{eq:PGPE} 
i\hbar\frac{\partial\psi(\mathbf{x})}{\partial t}
= {\cal P}\left\{\left(H_{\mathrm{sp}}+U|
\psi(\mathbf{x})|^2\right)\psi(\mathbf{x})\right\},
\end{equation} 
which is termed the projected Gross-Pitaevskii equation \cite{Blakie08}.  In the present work we use the rotationally invariant (Gauss-Laguerre) projector introduced in \cite{Bradley08}.  For a discussion of the evaluation of projections of the 2-body interaction and trap anisotropy operations in this basis, we refer the reader to \cite{Bradley08} and \cite{Wright08} respectively.  We simply note here that this method conserves both the energy and normalization of the field, and that the field angular momentum obeys the Ehrenfest relation of the continuous field theory
\begin{equation}\label{eq:Ehrenfest_Lz}
	\frac{d\overline{L_z}}{dt} = -\frac{i}{\hbar}\overline{L_zV(\mathbf{x})},
\end{equation}
[where the bar denotes spatial averaging: $\overline{A}=\int d \mathbf{x}\psi^*(\mathbf{x})A\psi(\mathbf{x})$], which is a consequence of the relation $[\mathcal{P},L_z]=0$. We note finally that as our formalism describes only the low-energy region $\mathbf{L}$, it does not account for any population above the cutoff.  The results of any classical-field theory are in general cutoff dependent, and the cutoff dependence of the formalism used here was quantified in \cite{Wright08}.  As noted in that work, ultimately this dependence can only be eliminated by the inclusion of above-cutoff physics, and currently no prescription for including such effects in general nonequilibrium scenarios is known.  In practice, however, the low-energy region defined by $E_R \sim 3\mu$, with $\mu$ the chemical potential (see Ref.~\cite{Wright08} and Refs. therein) is expected to include the most important dynamical mechanisms, the above-cutoff atoms influencing the dynamics only at a quantitative level.

\subsection{Simulation procedure}\label{subsec:Simulation_procedure}
Following \cite{Wright08}, we choose physical parameters corresponding to $^{23}\mathrm{Na}$ atoms confined in a strongly oblate trap, with trapping frequencies $(\omega_r,\omega_z)=2\pi\times(10,2000)$ rad/s.  The radial harmonic confinement defines the units of length ($r_0\equiv\sqrt{\hbar/m\omega_r}$) and time [one trap cycle (cyc) $\equiv2\pi\omega_r^{-1}$] in which we will quote our results.  The $s$-wave scattering length is $a=2.75nm$, placing our system in the quasi-two-dimensional regime (oscillator length $l_z = \sqrt{\hbar/m\omega_z} \gg a$), with an effective two-dimensional (2D) interaction parameter $U_{2\mathrm{D}} = 2\sqrt{2\pi}\hbar a/ml_z$.  We are thus justified in representing our system by a 2D model, provided that we maintain thermodynamic parameters (chemical potential $\mu$ and temperature $T$) such that $\mu+k_\mathrm{B}T\ll\hbar\omega_z$ \cite{Petrov00} at all times during the evolution.  The 2D representation is formalized in our classical-field method by choosing a cutoff $E_R=30\hbar\omega_r\ll\hbar\omega_z$, such that the low-energy space $\mathbf{L}$ excludes all modes with excitation along the $z$-axis.  In all our simulations this energy cutoff (projection operation) is effected in an inertial (laboratory) frame, and the low-energy space consists of $465$ single-particle modes.  

We form a finite-temperature initial state following the procedure of \cite{Wright09}:  we construct a (nonequilibrium) randomized classical-field configuration over the modes comprising the space $\mathbf{L}$, with chosen normalization, energy and angular momentum first integrals, and we evolve this state for some time ($10^4$ trap cycles), so that the field has time to migrate to an equilibrium configuration.  Following \cite{Wright09}, we take as the starting point for forming these configurations the ground Gross-Pitaevskii (GP) eigenstate with $N_0=1.072\times10^4$ atoms in a frame rotating at angular velocity $\Omega_0=0.35\omega_r$.  This state has chemical potential $\mu_\mathrm{g}=10.35\hbar\omega_r$ and energy $E_\mathrm{g} = 7.646\times10^4\hbar\omega_r$ in an inertial frame, and angular momentum $L_0=\hbar N_0$.  To this state we add energy and angular momentum, forming a configuration with $E=1.10E_\mathrm{g}$, and $L=1.20L_0$.  We find that the corresponding equilibrium state is one in which a single vortex precesses very close to the trap axis, i.e., with precession radius $r_\mathrm{v}\lesssim\eta_0$, where the healing length $\eta_0=0.20r_0$ is estimated from the density of the ground state \cite{Wright09}.  In this configuration the vortex is at equilibrium with the thermal component of the field, and the angular velocity $\Omega_\mathrm{c}$ of the condensate (which is the same as the cloud's at equilibrium) is close to that at the $\ell\equiv\langle L_z\rangle/\hbar N\rightarrow1^-$ limit of the mechanically unstable \cite{Butts99,Komineas05} precessing-vortex branch \cite{Papanicolaou05}.  If additional angular momentum is added to this state, the condensate mode becomes unstable to the nucleation of a second vortex, i.e., the single-vortex state we consider is essentially saturated with angular momentum for the given values of the other conserved integrals (normalization and energy).  We find for this equilibrium initial state $\Omega_\mathrm{c}\approx0.23\omega_r$, $\mu\approx12\hbar\omega_r$, $k_\mathrm{B}T\approx14\hbar\omega_r$ and condensate fraction $f_\mathrm{c}\approx0.91$. 
Having formed this initial state, we evolve field trajectories in the laboratory frame in the presence of trap anisotropies in the range $0.005\leq\epsilon\leq0.1$. This procedure corresponds in each case to the sudden introduction of the anisotropy at time $t=0$.  We evolve each of these trajectories with an adaptive integrator \cite{Davis_DPhil}, with accuracy chosen such that the relative change in field normalization is $\leq 10^{-9}$ per time step taken, until the vortex is expelled from the condensate and the total angular momentum of the field has been lost through its interaction with the anisotropy.

\section{Results}\label{sec:Results}
\begin{figure*}
	\includegraphics[width=0.9\textwidth]{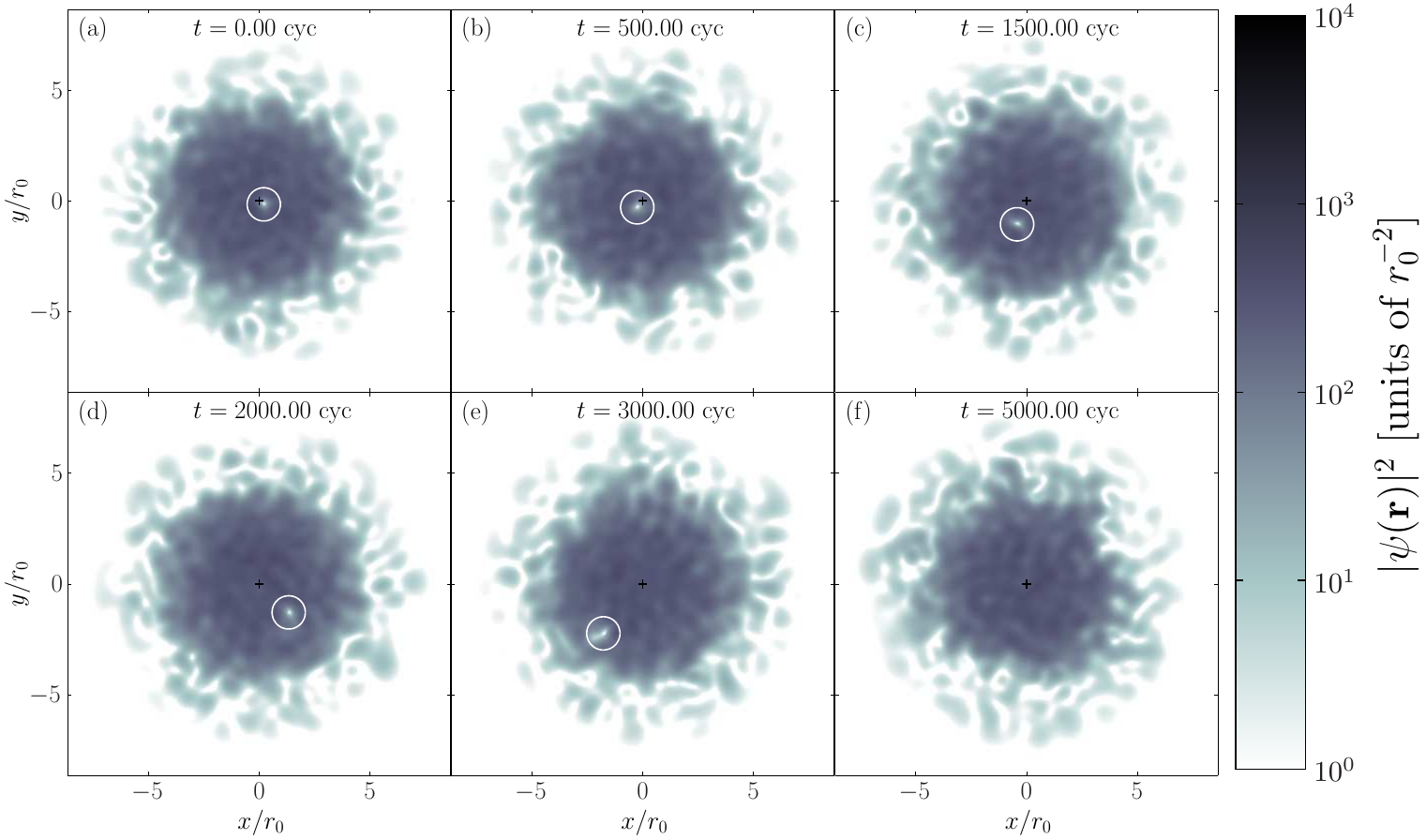}
	\caption{\label{fig:density_plots1} (Color online) (a-f) Classical-field densities at various times during the system evolution.  The white circle indicates the vortex position, and $+$ marks the coordinate origin (trap axis).  Parameters of the simulation are given in the text.}
\end{figure*}

In this section we consider the results of a classical-field simulation with a particular choice of the trap anisotropy, $\epsilon=0.025$.  The relaxation dynamics of the classical field in this case exhibit many features of interest, and serve as a point of comparison for simulations with weaker or stronger anisotropies.  Position-space densities of the classical field at representative times are presented in Fig.~\ref{fig:density_plots1}(a-f). 
Initially the vortex precesses very close to the trap axis [Fig.~\ref{fig:density_plots1}(a)], and it remains close to the trap axis for some time. Fig.~\ref{fig:density_plots1}(b) shows the density at time $t=500$ cyc, in which the vortex core remains within $\sim\eta_0$ of the trap axis.  By  $t\approx 1000$ cyc the vortex has begun to spiral out of the central density bulk which contains the condensate.  Fig.~\ref{fig:density_plots1}(c) shows that by $t=1500$ cyc the vortex has undergone significant radial displacement.  This increase in radial displacement continues [Fig.~\ref{fig:density_plots1}(d)], until the vortex core approaches the violently evolving condensate boundary [Fig.~\ref{fig:density_plots1}(e)]. At $t\approx3070$ cyc the vortex is lost into the peripheral thermal material, leaving the condensate vortex-free and essentially irrotational [Fig.~\ref{fig:density_plots1}(f)]. 
\subsection{Vortex trajectory}\label{subsec:vortex_trajectory}
\begin{figure}
	\includegraphics[width=0.45\textwidth]{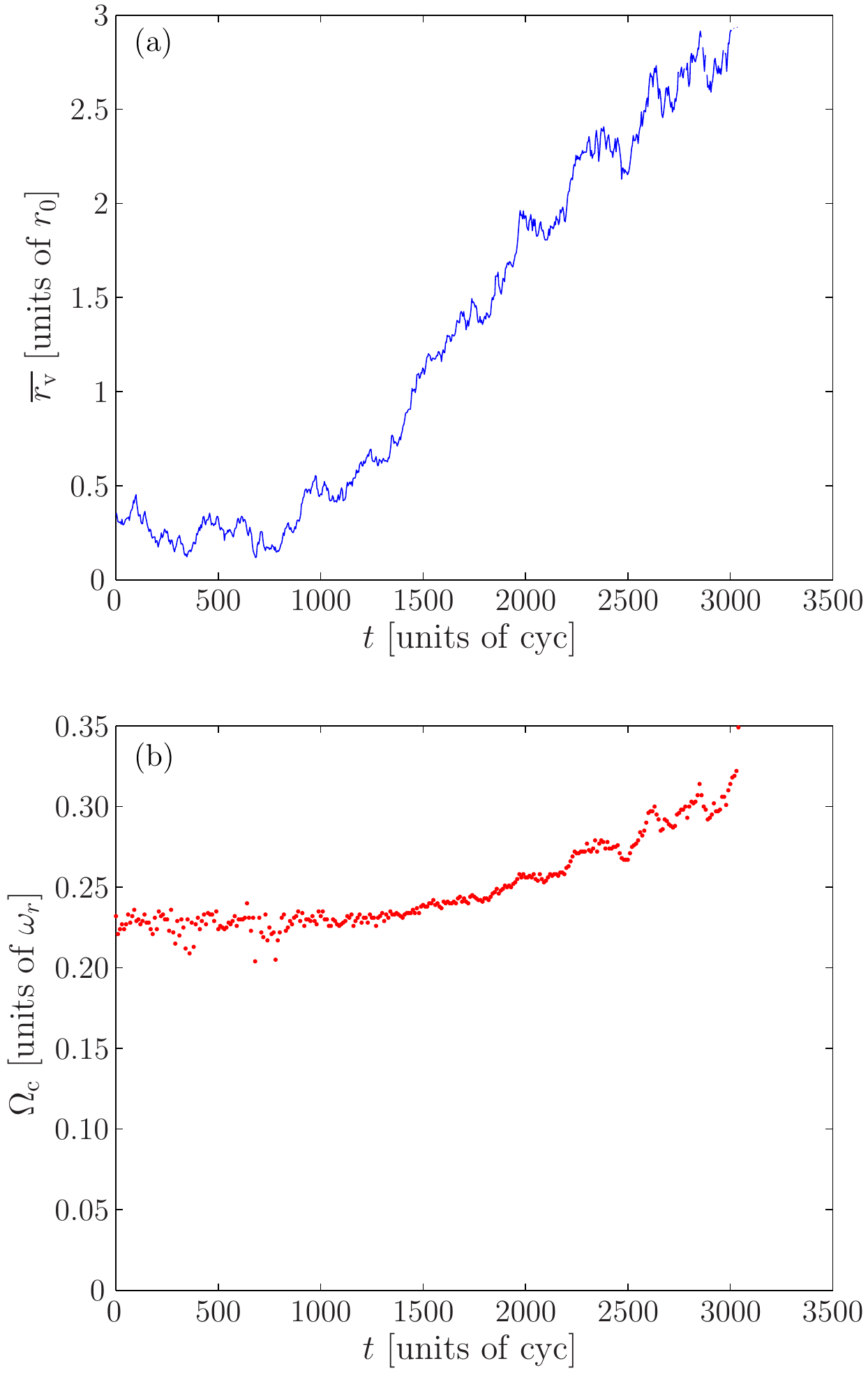}
	\caption{\label{fig:vortex_rad_freq} (Color online) Quantities characterizing the vortex motion.  (a) Vortex radial displacement and (b) angular velocity of the condensate orbital.}
\end{figure}

We now quantify the behavior of the vortex during its decay.  We observe that the vortex executes an essentially spiral-like motion as it decays, in agreement with the predictions of \cite{Fedichev99} and the simulations of \cite{Schmidt03,Jackson09}.  However, the vortex trajectory here is strongly stochastic, and the vortex orbits the trap axis many ($\sim 700$) times during its decay, so we do not present the full vortex trajectory.  To characterize the radial drift of the vortex, we track its location at a frequency of 25 samples per trap cycle, and average the resulting series over intervals of 100 samples ($4$ cyc) in order to smooth out rapid fluctuations due to thermal density fluctuations in the background field against which the vortex moves and uncertainties introduced by sampling the vortex location on a Cartesian grid.  In  Fig.~\ref{fig:vortex_rad_freq}(a) we present the smoothed vortex radius ($\overline{r_\mathrm{v}}$) data.
It is clear from Fig.~\ref{fig:vortex_rad_freq}(a) that there is an initial lag of $\sim1000$ cyc between the introduction of the trap anisotropy at $t=0$ cyc and the beginning of the upward trend in vortex displacement.  We note that the displacement radius does not increase monotonically, but exhibits large oscillations during its increase.  The period of these oscillations is often $\gtrsim100$ trap cycles, spanning many periods of the vortex orbit, in contrast to the trajectories presented in \cite{Schmidt03,Jackson09}, in which the vortex generally precesses at a greater radius on each passage about the trap center.  In Fig.~\ref{fig:vortex_rad_freq}(b) we plot the vortex precession frequency.  In practice, to evaluate this frequency, we follow the procedure of \cite{Wright09}, forming the covariance matrix (classical one-body density matrix)  
\begin{equation}\label{eq:density_matrix}
	\rho(\mathbf{x},\mathbf{x}') = \langle \psi^*(\mathbf{x}) \psi(\mathbf{x}') \rangle_\Omega, 
\end{equation}
where $\langle\cdots\rangle_\Omega$ denotes a time average, in a frame rotating at angular velocity $\Omega$ about the trap axis.  We vary $\Omega$ such that the largest eigenvalue of $\rho(\mathbf{x},\mathbf{x}')$ is maximized.  The value of $\Omega$ at which the maximum occurs is thus that of the rotating frame in which the coherent fraction of the classical field is most stationary, which forms a best estimate for the condensate angular frequency (vortex precession frequency) $\Omega_\mathrm{c}$ \cite{Wright09}.  
As we consider a nonequilibrium scenario, we construct Eq.~(\ref{eq:density_matrix}) by averaging over short time periods \cite{Blakie05}, in each case calculating the average of 250 consecutive classical-field samples over a ten cycle period.  
Fig.~\ref{fig:vortex_rad_freq}(b) shows that the vortex precession frequency determined in this manner increases as the vortex displacement increases, as is well known for a zero-temperature condensate \cite{Fetter01,Komineas05} and observed in the simulations of \cite{Schmidt03,Jackson09}.  Moreover, we observe that the oscillations in vortex radius $r_\mathrm{v}$ at late times $t\gtrsim2000$ cyc are accompanied by oscillations in the condensate angular velocity. The oscillations in the two are positively correlated, i.e., the decreases in vortex radius are associated with periods of slowing of the condensate rotation during its otherwise steady increase with time.
\subsection{Rotational dynamics of condensate and thermal cloud}\label{subsec:rotational_dynamics}
\begin{figure}
	\includegraphics[width=0.45\textwidth]{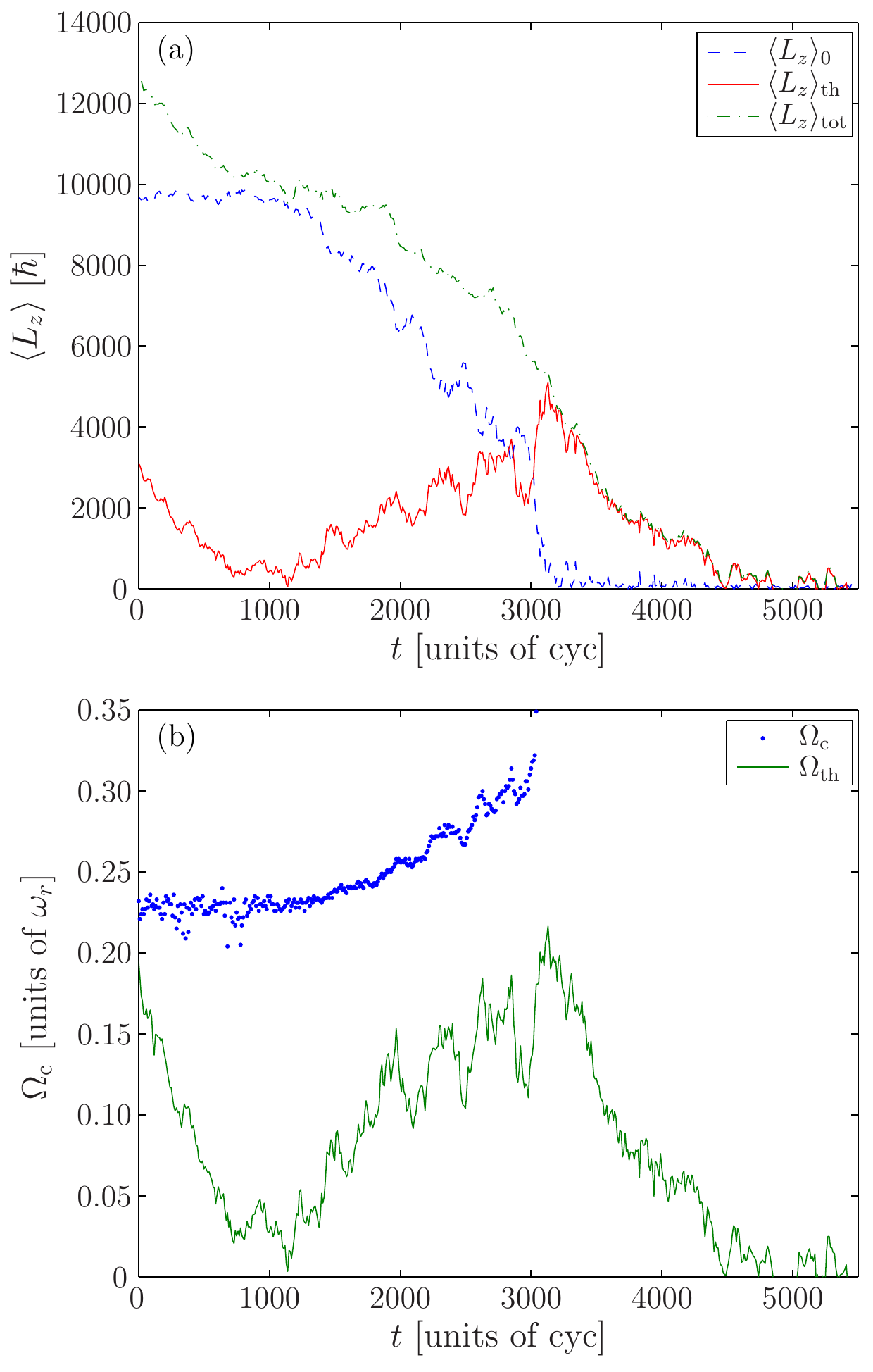}
	\caption{\label{fig:ang_mom_and_vel} (Color online) Quantities characterizing the temporal evolution of the condensate and thermal cloud.  (a) Angular momenta of the condensate and thermal cloud, and total angular momentum of the field. (b) Angular velocities of the condensate and thermal cloud.}
\end{figure}

The above definition of the condensate mode in terms of short-time covariance matrix eigenvectors allows us to resolve the dynamics of the condensed and noncondensed components of the field.  As in \cite{Wright09}, we introduce the decomposition of the classical-field one-body density matrix in terms of its eigenvectors $|\chi_i\rangle$ and corresponding eigenvalues $n_i$ (indexed in order of decreasing eigenvalue),
\begin{equation}
	\rho = n_0|\chi_0\rangle\langle \chi_0| + \sum_{k\geq0} n_k |\chi_k\rangle\langle \chi_k| \equiv \rho_0 + \rho_\mathrm{th},
\end{equation}
which separates it into condensed and noncondensed parts.
As in \cite{Wright09}, we then define the averages of a single-body operator $J$ in the condensed and noncondensed components of the field by $\langle J\rangle_0 = \mathrm{Tr}\{\rho_0J\}$ and $\langle J\rangle_\mathrm{th} = \mathrm{Tr}\{\rho_\mathrm{th}J\}$ respectively.
We use this decomposition to calculate the angular momentum of the condensate  ($\langle L_z \rangle_0$) and thermal cloud ($\langle L_z \rangle_\mathrm{th}$).  These quantities are presented in Fig.~\ref{fig:ang_mom_and_vel}(a), along with the total angular momentum of the field ($\langle L_z\rangle_\mathrm{tot}$).
We observe that in the first few hundred trap cycles after the introduction of the trap anisotropy, the angular momentum of the thermal cloud [solid line in Fig.~\ref{fig:ang_mom_and_vel}(a)] undergoes approximately exponential decay, reaching values close to zero by $t\approx1000$ trap cycles. By contrast, the angular momentum of the condensate (dashed line) is essentially unchanged during this period, which we identify with the initial period of quiescence of the vortex noted in Sec.~\ref{subsec:vortex_trajectory}.  At $t\approx1100$ cyc, the angular momentum of the condensate begins to decay, and after $t\approx3200$ cyc the condensate is essentially irrotational, with some small value of angular momentum remaining in the form of surface excitations.  We note that the angular momentum of the thermal field component \emph{increases} over the period $t\in[1000,3200]$ cyc.  
We understand this as follows: Initially the angular momentum of the thermal cloud is lost due to its interaction with the trap anisotropy.  This makes the vortex thermodynamically unstable, and thus it begins to decay, liberating its angular momentum to the cloud.  The cloud loses its angular momentum to the trap at a rate proportional to its angular velocity (frictional loss $d\langle L_z\rangle/dt\propto -\Omega_\mathrm{th}$ \cite{Zhuravlev01}), which is initially small.  Therefore, the cloud's angular momentum increases as it gains angular momentum from the vortex faster than it loses it to the trap.  As the cloud's rotation rate increases the rate of dissipation of $\langle L_z \rangle_\mathrm{th}$ increases, however, the precessing vortex condensate exhibits a thermodynamic anomaly: as the condensate loses angular momentum, its angular velocity \emph{increases} \cite{Butts99,Papanicolaou05,Komineas05}.  The rate at which the angular momentum of the vortex is dissipated to the thermal cloud is proportional to their relative velocities \cite{Fedichev99}, and so in the present case this transfer dominates the loss rate of angular momentum of the thermal cloud throughout the vortex decay.  As shown in Fig.~\ref{fig:ang_mom_and_vel}(b), this leads to `run-away' spin-up of the vortex precession frequency, with the thermal-cloud rotation rate being driven up in sympathy.

Closer inspection of the angular momenta $\langle L_z\rangle_0$ and $\langle L_z\rangle_\mathrm{th}$ reveals negatively correlated oscillations in the angular momenta of the two components during the decay, as a result of the detailed nonlinear dynamics of angular momentum exchange and loss during the decay process.  We identify this oscillation in angular momentum transfer with the oscillations in vortex rotation rate discussed in Sec.~\ref{subsec:vortex_trajectory}.
In Fig.~\ref{fig:ang_mom_and_vel}(b) we plot the angular velocity of the condensate $\Omega_\mathrm{c}$ (dots) derived in the estimation of the condensate fraction (Sec.~\ref{subsec:vortex_trajectory}), and the angular velocity of the thermal cloud $\Omega_\mathrm{th} = \langle L_z \rangle_\mathrm{th} / \langle \Theta_\mathrm{c} \rangle_\mathrm{th}$ (line), where we assume the expectation value of the classical moment of inertia $\Theta_\mathrm{c}\equiv r^2$ as an estimate of the cloud's true moment of inertia.  In \cite{Wright09} we discussed the level of uncertainty inherent in this procedure, nevertheless it yields a clear qualitative description of the decay dynamics.  The oscillations in rotation rate of the thermal cloud are clearly visible here, and we note that they are \emph{positively} correlated with the oscillations in angular velocity of the condensate, due to its anomalous rotational response.   We note that oscillatory behavior arises already in a linear analysis of the arrest of a rotating Boltzmann gas \cite{Guery-Odelin00} by a trap anisotropy.  It is therefore not surprising that similar oscillations occur in the transfer of angular momentum from the condensate to the nonequilibrium thermal field, considering the complexity of their coupled dynamics (cf. \cite{Zhuravlev01}), and the anomalous response of the vortical condensate.

Finally, we note that the cloud angular momentum (rotation rate) reaches its peak when the vortex leaves the condensate ($t\approx3200$ cyc), after which it undergoes a second near-exponential decay phase, the angular momentum of the field decaying such that by $t\approx5000$ cyc, only the thermodynamic fluctuations in $\langle L_z \rangle_\mathrm{th}$ exhibited by the finite-temperature field at rest in the anisotropic potential remain.

\subsection{Heating of the atomic field}\label{subsec:heating}
We now consider the heating of the atomic field during the arrest of its rotation.  As the system we evolve is Hamiltonian, with a time-independent potential, the total energy of the classical field is a constant of the motion.  Consequently, the trap anisotropy dissipates the angular momentum of the field by converting the rotational kinetic energy of the field into internal energy \cite{Zhuravlev01}, and we therefore expect some heating of the field to occur, due to this redistribution of energy.  We can estimate the heating of the field as follows:  The rotational energy of the gas is initially $(E_\mathrm{rot})_\mathrm{i}=\Omega_\mathrm{i}L_\mathrm{i}\approx2.7\times10^3\hbar\omega_r$.  We compare this to the initial thermal energy, which should be reasonably well estimated by the energy added to the ground vortex state in forming the initial thermal state (Sec.~\ref{subsec:Simulation_procedure}), $E_\mathrm{th}\sim0.1E_\mathrm{g}\approx7.6\times10^3\hbar\omega_r$.  Assuming a linear relationship between thermal energy and temperature (which should be valid for the low temperatures we consider here, see \cite{Davis05}), we might therefore expect an increase in the field temperature of $\sim30\%$ during the arrest of the condensate rotation.  

\begin{figure}
	\includegraphics[width=0.45\textwidth]{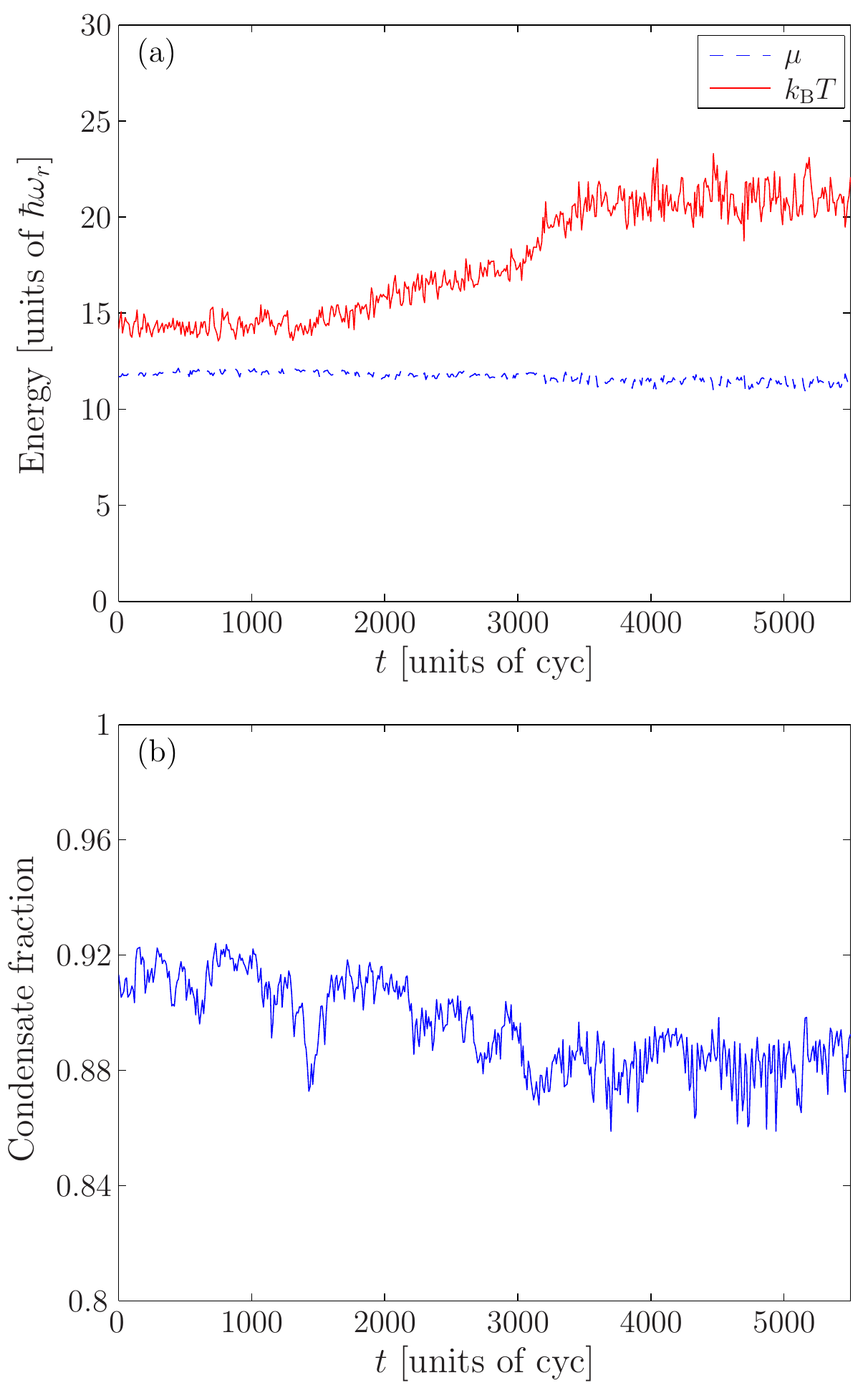}
	\caption{\label{fig:cfrac_and_temp} (Color online) Heating of the field.  Evolution of (a) the effective chemical potential and temperature and (b) the condensate fraction.} 
\end{figure}

In order to quantify the heating and its development during the field evolution, we estimate the (effective) thermodynamic parameters of the field (chemical potential $\mu$ and temperature $T$) using the procedure of \cite{Wright08,Wright09}.
As in \cite{Wright09}, we allow for the differential rotation between the thermal cloud and the frame in which the classical-field cutoff is effected (the laboratory frame, in the present case). We perform the fit to the wing of the (time-averaged) radial distribution $n(r)$ of the classical field, i.e., we fit over the radii range $r\in[r_-+0.5r_0,r_\mathrm{tp}]$ where $r_-$ marks the minimum of the Hartree-Fock effective potential $V_\mathrm{eff}=(m/2)[\omega_r^2-\Omega_\mathrm{th}^2]r^2 + 2U_\mathrm{2D} n(r)$ \cite{NoteA}, and $r_\mathrm{tp}$ is the semiclassical turning-point of the low-energy space $\mathbf{L}$ \cite{Wright09}.  We fit to radial densities of the classical field averaged over $10$ cyc periods, and assume the values of $\Omega_\mathrm{th}$ calculated in Sec.~\ref{subsec:rotational_dynamics}.  The resulting estimates for $\mu$ and $T$ are presented in Fig.~\ref{fig:cfrac_and_temp}(a).
We find that the temperature (dashed line) is approximately constant during the first $\sim1000$ cycles of the field evolution, corresponding to the initial arrest of the thermal field component.  The initial angular momentum of the thermal component is small, comprising some $\sim25\%$ of the total angular momentum of the field, and so any heating of the field due to the redistribution of the associated rotational energy is possibly too small to resolve above the uncertainty in the temperature estimates. Beginning at $t\approx1100$ cyc, corresponding to the start of the vortex-decay phase, the temperature exhibits a steady, approximately linear increase.  We associate this increase with the conversion of rotational kinetic energy to thermal energy, due to the action of the trap anisotropy, and presumably also as a result of the scattering of excitations by the vortex which produces the frictional effect \cite{Fedichev99}.  There is a final sharp rise in temperature at $t\approx3000-3200$ trap cycles.  This rise is perhaps due to the final decay of the vortex into excitations at the surface of the condensate \cite{Fedichev99}.  We note however that the field, and in particular the condensate surface, undergoes strong fluctuations as the system passes through this transition to the vortex-free state, with, e.g., multiple (ghost) vortices present at the condensate surface at times.  This can be viewed as the system essentially `reversing' through the surface-wave instability arising from the relative motion of condensate and thermal cloud, in which vortices spontaneously grow from surface excitations \cite{Williams2002a, Penckwitt02, Wright08}.  The thermodynamic parameters may therefore be ill-defined during this period.  After this period of strong surface fluctuations, the temperature levels off at $T\approx 21 \hbar\omega_r/k_\mathrm{B}$, corresponding to heating of $\approx50\%$ during the arrest.  We note that the chemical potential [solid line in Fig.~\ref{fig:cfrac_and_temp}(a)] exhibits a slight downward trend beginning around $t\approx1100$ cyc, falling by $\sim0.4\hbar\omega_r$, though we note that this change is of the same order as the variation in estimates for $\mu$ obtained at late times $t\gtrsim3500$ cyc. 

In Fig.~\ref{fig:cfrac_and_temp}(b) we plot the condensate fraction $f_\mathrm{c} \equiv n_0 / \sum_{k\geq0}n_k$ obtained from the procedure outline in Sec.~\ref{subsec:vortex_trajectory}.  The fluctuations in estimates of this quantity are large, as is expected given the short time scale ($10$ cyc) over which the appropriate averages are taken, and the nonequilibrium nature of the field.  However the condensate fraction estimates exhibit a clear downwards trend as time proceeds, and the condensate fraction drops by $\sim3\%$ during the arrest.

\section{Dependence on trap anisotropy}\label{sec:Anisotropy_dependence}
We now consider the effect of varying the trap anisotropy $\epsilon$ on the behavior of the classical-field trajectories.  We intuitively expect the rate at which the trap dissipates the angular momentum of the thermal cloud to depend strongly on the magnitude of the anisotropy, and a quantitative model for this dependence in the case of a classical (Boltzmann) gas was presented in \cite{Guery-Odelin00}.  By contrast, the efficiency with which the thermal cloud extracts angular momentum from the condensate is dictated by the (longitudinal) mutual-friction coefficient, which depends on the temperature and chemical potential of the field, in addition to its microscopic properties \cite{Hall56,Sonin97,Fedichev99}.  We therefore expect, as discussed in the vortex-continuum analysis of \cite{Zhuravlev01}, to explore different regimes of relaxation dynamics as we vary the trap anisotropy and consequently the relative strengths of vortex-cloud and cloud-trap friction.  In this section we consider the evolution of the classical field in the presence of anisotropies of different magnitudes, and observe the resulting differences in the behavior of the fields. 
\subsection{Weak damping}\label{subsec:weak_damping}
\begin{figure*}
	\includegraphics[width=0.9\textwidth]{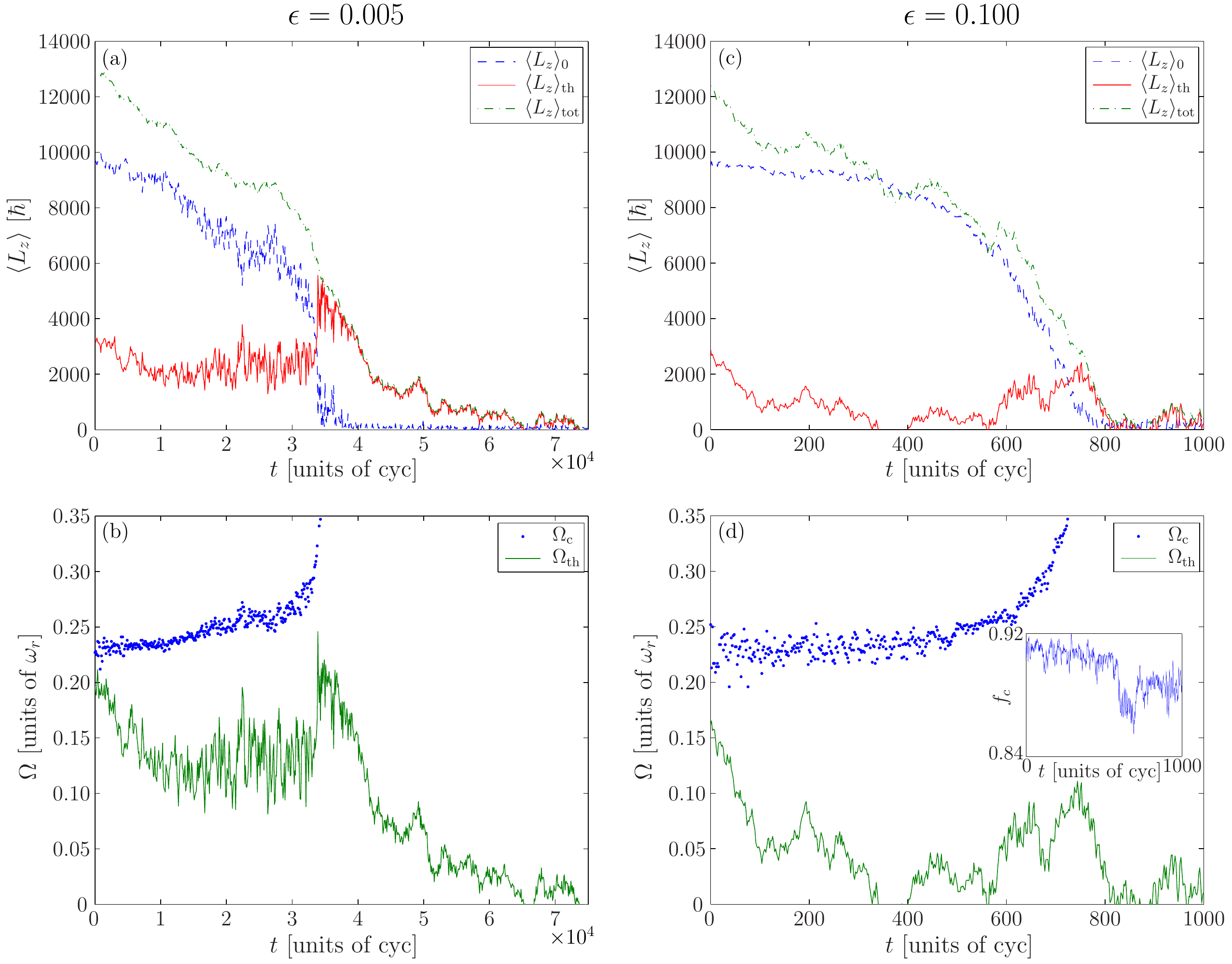}
	\caption{\label{fig:adiabatic_and_not} (Color online) (a) Angular momenta and (b) angular velocities of field components in a simulation with trap anisotropy $\epsilon=0.005$. (c) Angular momenta and (d) angular velocities of field components in a simulation with trap anisotropy $\epsilon=0.100$. Inset: Evolution of the condensate fraction in the case $\epsilon=0.100$.}
\end{figure*}

In Fig.~\ref{fig:adiabatic_and_not}(a) we plot the evolution of the angular momentum of the condensate and thermal cloud in the presence of a weak trap anisotropy ($\epsilon=0.005$).  
We observe that after an initial decline over the first $\approx10^4$ trap cycles, the cloud angular momentum, despite exhibiting large fluctuations, maintains a reasonably steady mean value which increases slowly over the period $t\sim[10^4,3\times10^4]$ trap cycles.  During this period the angular momentum of the condensate drops by approximately a factor of $2$, and large fluctuations in the distribution of angular momentum between the two components are visible.  This behavior is again apparent in the vortex precession frequency and cloud rotation rate presented in Fig.~\ref{fig:adiabatic_and_not}(b).  This shows that the cloud exhibits a steady-state nonequilibrium behavior, i.e., its rotation rate remains somewhere between that of the precessing condensate mode, and that of the trap ($\Omega_\mathrm{tr}=0$).  In this regime its angular momentum remains approximately constant, as the rate at which it gains angular momentum from the decaying vortex matches the rate at which it loses angular momentum to the trap.  We expect the angular velocity at which this balance occurs, and indeed whether such a regime occurs at all, to depend strongly on the relative strengths of the vortex-cloud and cloud-trap friction.  We note that the angular velocity of the cloud appears to slowly increase over time, as the condensate loses angular momentum and its angular velocity increases, shifting the cloud rotation rate at which the angular momentum transfer rates are balanced.  At $t\approx3.2\times10^4$ cyc the vortex's precession accelerates as it approaches the condensate boundary, and during the period $t\in[3.2\times10^4,3.4\times10^4]$ cyc the adiabaticity of the vortex dissipation is lost as the field enters the critical regime associated with nucleation of the vortex at the surface \cite{Wright09}.  In this regime the condensate surface is unstable and as noted in Sec.~\ref{subsec:heating}, multiple `ghost' vortices may be present simultaneously at the condensate boundary.  A large amount of angular momentum is transferred non-adiabatically to the thermal cloud during this period of criticality, while small surface-mode oscillations of the condensate persist until $t\approx4\times10^4$ cyc.  The angular momentum is subsequently dissipated from the thermal cloud over a period of $\sim3\times10^4$ cyc, in an oscillatory but approximately exponential decay phase.
\subsection{Strong damping}\label{subsec:strong_damping}
We now turn our attention to a scenario of vortex arrest due to the presence of a strong trap anisotropy ($\epsilon=0.1$).  The dynamics in this case are strongly nonequilibrium and the arrest of the field's rotation occurs on a shorter time scale than the cases already considered, and so here we form the density matrix Eq.~(\ref{eq:density_matrix}) by averaging classical-field samples over a shorter period of two trap cycles.  In this case the angular momentum of the thermal cloud [solid line in Fig.~\ref{fig:adiabatic_and_not}(c)] is rapidly depleted, and actually fluctuates below zero by $t\approx400$ trap cycles.  The condensate is slower to respond; during this period the condensate angular momentum decreases by $\sim7\%$.  Subsequently, the cloud angular momentum fluctuates strongly, and rises to $\sim2000\hbar$ (close to its initial value), as the vortex is rapidly expelled from the condensate.  The angular momentum is then dissipated from the cloud over a period $\lesssim200$ cyc, and thereafter the angular momentum of the field fluctuates about zero.  The rapid expulsion of the vortex, and strong fluctuations of the cloud rotation, are again visible in the calculated angular velocities [Fig.~\ref{fig:adiabatic_and_not}(d)].  

The behavior of the cloud angular momentum in this trajectory suggests that the cloud is overcritically damped by the trap anisotropy \cite{Guery-Odelin00}.  Its response to the anisotropy appears unhindered by its coupling to the condensate via the vortex core, and it quickly yields the angular momentum it acquires from the vortex to the trap, despite its small angular velocity $\Omega_\mathrm{th}\lesssim0.1$.  The relaxation process in this case is violently non-adiabatic, as evidenced by the strong fluctuations in the cloud angular momentum.  Indeed during the period $t\sim[600,700]$ cyc large, long-wavelength surface oscillations are visible in the field as the vortex precesses rapidly near the condensate boundary.  It appears that the strong trap anisotropy and rapid `stirring' motion of the vortex conspire to strongly perturb the condensate bulk in this regime.  During this period the measured condensate fraction is suppressed [inset to Fig.~\ref{fig:adiabatic_and_not}(d)] due to the strongly nonequilibrium behavior of the condensate, in which surface-wave excitations define frames of rotation distinct from that of the vortex.  The decomposition of the field into condensed and noncondensed components must therefore be viewed with some caution in this strongly nonequilibrium scenario.  Nevertheless, it is clear that the effect of the trap on the cloud dominates the vortex-cloud coupling in this scenario, in stark contrast to the near-adiabatic decay scenario of Sec.~\ref{subsec:weak_damping}.

\subsection{Decay times}\label{subsec:decay_times}
\begin{figure}
	\includegraphics[width=0.45\textwidth]{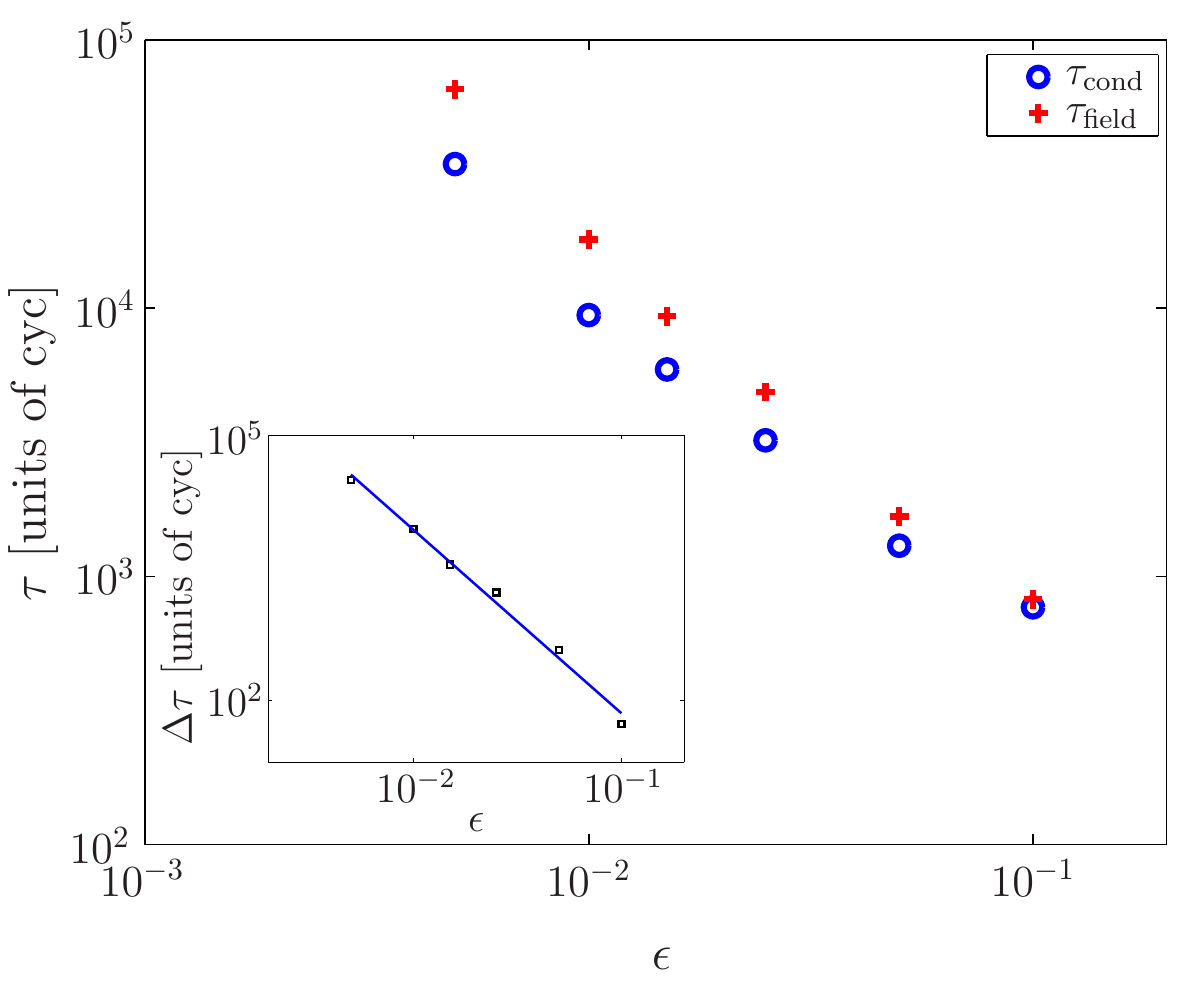}
	\caption{\label{fig:relaxation} (Color online) Relaxation of the field.  Circles (plusses) represent the times at which the vortex leaves the condensate (total field angular momentum reaches zero). Inset: Cloud spin-down times and the linear fit performed to extract their scaling with trap ellipticity.}
\end{figure}

We now consider the dependence of the relaxation times on the strength of the trap anisotropy.  As the relaxation of both the condensate and the thermal cloud is generally nonexponential, we consider the times at which the condensate becomes irrotational, and at which the total angular momentum of the field is lost.  Due to the persistence of surface excitations which prevent the condensate angular momentum from reaching zero, we define the condensate stopping time $\tau_\mathrm{cond}$ as that at which the angular momentum of the condensate first drops below $0.02\hbar$ per particle, and the field stopping time as the first \emph{subsequent} time $\tau_\mathrm{field}$ at which the cloud angular momentum reaches zero.  These times are presented in Fig.~\ref{fig:relaxation}.  
We observe that the condensate arrest times (circles in Fig.~\ref{fig:relaxation}) vary by two orders of magnitude over the range of trap anisotropies we consider.  Moreover, the total-field relaxation times (plusses) become increasingly longer than the vortex relaxation times as the trap anisotropy is weakened, causing slower dissipation of the cloud angular momentum even in the absence of the vortex.  We therefore consider the time $\Delta \tau = \tau_\mathrm{field}-\tau_\mathrm{cond}$ over which the angular momentum of the cloud dissipates following the expulsion of the vortex.  Although the angular momentum lost in this final damping phase varies over the range $\langle L_z \rangle_\mathrm{th} \sim 3000-5000\hbar$, precluding a precise analysis, the times $\Delta \tau$ provide a useful characterization of the dependence of the cloud relaxation on the trap anisotropy. We perform a linear fit to $\Delta\tau$ as a function of $\epsilon$ in log-log space (inset to Fig.~\ref{fig:relaxation}), and find the scaling $\Delta \tau \propto \epsilon^{-2.1}$, in good agreement with the scaling $t_\mathrm{down}\propto \epsilon^{-2}$ for thermal-cloud spin-down predicted for weak anisotropies by Gu\'ery-Odelin \cite{Guery-Odelin00}. 

\section{Summary and Conclusions}\label{sec:Conclusions}
We have carried out the first simulations of the arrest of a rotating Bose-Einstein condensate due to the presence of a trap anisotropy which includes the coupled nonequilibrium dynamics of the condensate and thermal cloud.  Our method makes no assumptions of stationarity of a thermal bath, nor are the two components artificially given disparate rotational parameters.  Rather, our approach describes the dynamical migration of an \emph{equilibrium} rotating thermal state to a new, irrotational equilibrium, solely due to the action of the trapping potential.

We observe for all parameters we considered that the rotation rates of the condensate and thermal component are distinct during the decay.  The anomalous rotational response of the precessing-vortex condensate can lead to a counter-intuitive spin-up of the thermal component during the decay, and we observe nonequilibrium oscillations in transfer of angular momentum between the two components as the vortex responds to the dissipative effect of the thermal cloud, which is itself damped by the trapping potential anisotropy.

For trap anisotropies that are weak, the thermal field settles to a nonequilibrium steady state, with rotation rate intermediate between that of the condensate and that of the (static) trap.  In this scenario the angular momentum of the condensate is slowly depleted while that of the thermal cloud remains nearly constant, until the vortex nears the condensate boundary and the linearity of the vortex decay breaks down.  For stronger trap anisotropies, the angular momentum of the thermal cloud may be almost entirely depleted before the condensate responds. 

We quantified the heating of the atomic field during the arrest, and found it to be commensurate with the conversion of rotational kinetic energy into thermal energy by the trap anisotropy.  We also considered the time scales over which the condensate angular momentum and total field angular momentum were dissipated, and found reasonable agreement between the scaling of the thermal cloud spin-down time with trap anisotropy and the predictions of a Boltzmann gas model \cite{Guery-Odelin00}. 

Possible extensions of this work include the description of vortex-lattice arrest, and the effects of temperature on the decay dynamics.  In particular it would be interesting to study how the nonequilibrium rotational dynamics of the condensate approach those of a rigid body as the vortex density increases, and the effect of temperature on the cloud rotation in the steady-state scenario.
\begin{acknowledgments}
We wish to acknowledge valuable discussions with A. Polkovnikov.  
This work was supported by the New Zealand Foundation for Research, Science and Technology under Contract No. NERF-UOOX0703 and No. UOOX0801.
\end{acknowledgments}
\bibliographystyle{prsty}

\begin{thebibliography}{10}
\bibitem{NoteA}
{We refer here to the local minimum associated with the boundary of the
  condensate, rather than any minimum that may occur due to the presence of the
  vortex core.}
\end{thebibliography}

\end{document}